\documentclass[a4paper, cleveref, autoref, thm-restate]{lipics-v2021}

\hideLIPIcs  

\title{Dependency Tuples for Almost-Sure Innermost Termination of Probabilistic Term
  Rewriting\\(Short WST Version)} 

\titlerunning{Dependency Tuples for Innermost AST of PTRSs} 

\author{Jan-Christoph Kassing}{\and \url{https://verify.rwth-aachen.de/jckassing/}}{kassing@cs.rwth-aachen.de}{https://orcid.org/0009-0001-9972-2470}{}

\author{Jürgen Giesl}{LuFG Informatik 2, RWTH Aachen University, Aachen, Germany \and \url{https://verify.rwth-aachen.de/giesl/}}{giesl@informatik.rwth-aachen.de}{https://orcid.org/0000-0003-0283-8520}{}

\authorrunning{J.-C.\ Kassing and J.\ Giesl}

\Copyright{Jan-Christoph Kassing and Jürgen Giesl}

\relatedversiondetails{See \cite{report}. Full version, including all
  proofs}{https://arxiv.org/abs/2305.11741}



\funding{funded by the Deutsche Forschungsgemeinschaft (DFG, German Research Foundation) - 235950644 (Project GI 274/6-2) and DFG Research Training Group 2236 UnRAVeL}

\nolinenumbers 

\makeatletter \newcommand*\bigcdot{\mathpalette\bigcdot@{.4}}
\newcommand*\bigcdot@[2]{\mathbin{\vcenter{\hbox{\scalebox{#2}{$\m@th#1\bullet$}}}}}
\makeatother

\renewcommand{\emptyset}{\varnothing}

\newcommand{\disabledcomment}[1]{}
\newcommand{\oldcomment}[1]{}

\newcommand{\dontprint}[1]{}

\renewcommand{\epsilon}{\varepsilon}

\newcommand{\IN}{\mathbb{N}}

\newcommand{\R}{\mathcal{R}}

\newcommand{\PP}{\mathcal{P}}
\newcommand{\SSS}{\mathcal{S}}

\newcommand{\aprove}{\textsf{AProVE}}

\crefname{definition}{Def.}{Def.}
\crefname{example}{Ex.}{Ex.}
\crefname{counterexample}{Counterex.}{Counterex.}
\crefname{appendix}{App.}{App.}
\crefname{ex}{Ex.}{Ex.}
\crefname{theorem}{Thm.}{Thm.}
\crefname{lemma}{Lemma}{Lemmas}
\crefname{remark}{Rem.}{Rem.}
\crefname{section}{Sect.}{Sect.}
\crefname{subsection}{Sect.}{Sect.}
\crefname{subsubsection}{Sect.}{Sect.}
\crefname{line}{Line}{Lines}
\crefname{corollary}{Cor.}{Cor.}
\crefname{figure}{Fig.}{Fig.}
\crefname{enumi}{}{}
\crefname{algorithm}{Alg.}{Alg.}

\newcommand{\ts}{\mathsf{s}}
\renewcommand{\O}{\mathcal{O}}
\newcommand{\tf}{\mathsf{f}}
\newcommand{\tg}{\mathsf{g}}
\renewcommand{\th}{\mathsf{h}}
\newcommand{\ta}{\mathsf{a}}
\newcommand{\tb}{\mathsf{b}}

\newcommand{\tminus}{\mathsf{minus}}
\newcommand{\tdiv}{\mathsf{div}}

\newcommand{\trw}{\mathsf{rw}}

\newcommand{\tM}{\mathsf{M}}
\newcommand{\tD}{\mathsf{D}}
\newcommand{\tF}{\mathsf{F}}
\newcommand{\tG}{\mathsf{G}}
\newcommand{\tH}{\mathsf{H}}
\newcommand{\tA}{\mathsf{A}}
\newcommand{\tB}{\mathsf{B}}

\newcommand{\tqs}{\mathsf{qs}}

\newcommand{\tstop}{\mathsf{stop}}

\newcommand{\F}[1]{\mathfrak{#1}}

\newcommand{\ctleaf}{\operatorname{Leaf}}

\newcommand{\Pol}{\operatorname{Pol}}
\newcommand{\Proc}{\operatorname{Proc}}

\newcommand{\Com}[1]{\mathsf{c}_{#1}}

\newcommand{\TSet}[2]{\mathcal{T}\left(#1,#2\right)}

\newcommand{\VSet}{\mathcal{V}}

\newcommand{\DTuple}[1]{\mathcal{DT}(#1)}

\newcommand{\FDist}{\operatorname{FDist}}
\newcommand{\Supp}{\operatorname{Supp}}

\newcommand\nameft\textrm

\makeatletter
\newcommand{\oset}[2]{%
  {\mathop{#2}\limits^{\vbox to 1\ex@{\kern-\tw@\ex@
   \hbox{\scriptsize #1}\vss}}}}
\makeatother

\makeatletter
\newcommand{\osetthree}[2]{%
  {\mathop{#2}\limits^{\vbox to 3\ex@{\kern-\tw@\ex@
   \hbox{\scriptsize #1}\vss}}}}
\makeatother

\makeatletter
\newcommand{\osetfive}[2]{%
  {\mathop{#2}\limits^{\vbox to 5\ex@{\kern-\tw@\ex@
   \hbox{\scriptsize #1}\vss}}}}
\makeatother

\makeatletter
\newcommand{\osetminus}[2]{%
  {\mathop{#2}\limits^{\vbox to -2\ex@{\kern-\tw@\ex@
   \hbox{\scriptsize #1}\vss}}}}
\makeatother

\newcommand{\itor}{\mathrel{\smash{\stackrel{\raisebox{3.4pt}{\scriptsize $\mathsf{i}\:$}}%
{\smash{\rightarrow}}}_{\R}}}

\newcommand{\nonprob}{\mathrm{np}}

\newcommand{\itononprobsstar}{\mathrel{\smash{\stackrel{\raisebox{3pt}{\scriptsize $\mathsf{i}$}}%
{\smash{\rightarrow}}}_{\normalfont{\nonprob}(\SSS)}^*}}

\usepackage{tikz}
\usetikzlibrary{arrows}
\usetikzlibrary{shapes.geometric}
\usetikzlibrary{arrows.meta}
\usepackage{nicefrac,xfrac}
\usepackage{mathtools}
\usepackage{stmaryrd}
\usepackage{amsmath,amssymb}
\usepackage{wrapfig}

\begin{document}

\maketitle

\begin{abstract}
    Dependency pairs are one of the most powerful techniques to analyze termination of term rewrite systems (TRSs) automatically.
    We adapt the dependency pair framework to the probabilistic setting in order to prove almost-sure innermost termination of probabilistic TRSs.
    To evaluate its power, we implemented the new framework in our tool \aprove.
\end{abstract}

\section{Introduction}\label{sec-introduction}
Techniques and tools to analyze innermost termination of TRSs automatically are successfully used for termination analysis of programs in many languages.
While there exist several classical orderings for proving termination of TRSs, a \emph{direct} application of these orderings is usually too weak for TRSs that result from actual programs. However, these orderings can be used successfully within the \emph{dependency pair} (DP) framework \cite{arts2000termination,giesl2006mechanizing}, which allows for modular termination proofs and is one of the most powerful techniques for termination analysis of TRSs that is used in essentially all current termination tools for TRSs.

On the other hand, \emph{probabilistic} programs are used to describe randomized algorithms and probability distributions. 
To use TRSs also for such programs, \emph{probabilistic term rewrite systems} (PTRSs) were introduced in \cite{bournez2005proving}. 
A probabilistic program is \emph{almost-surely terminating} (AST) if the probability for termination is $1$.
While there exist automatic approaches to prove AST for probabilistic programs on numbers, this is the first approach to prove AST for PTRSs.
The only other related tool
 was presented in~\cite{avanzini2020probabilistic}, where orderings
based on interpretations were adapted to prove \emph{positive almost-sure termination} of PTRSs,
i.e., that the expected number of rewrite steps is finite.

In this paper, we adapt DPs to the probabilistic setting and present the first DP framework for probabilistic term rewriting. 
We also present an adaption of the technique from~\cite{avanzini2020probabilistic} for the direct application of polynomial interpretations in order to prove AST of PTRSs.

\section{Probabilistic Term Rewriting}

We assume familiarity with term rewriting \cite{baader_nipkow_1999} and the DP framework \cite{arts2000termination,giesl2006mechanizing}.
In this paper, we restrict ourselves to \emph{innermost} rewriting.
In contrast to TRSs, a PTRS \cite{avanzini2020probabilistic,bournez2005proving} has finite multi-distributions on the right-hand side of rewrite rules.
A finite \emph{multi-distribution} $\mu$ on a set $A \neq \emptyset$ is a finite multiset
of pairs $(p:a)$, where $0 < p \leq 1$ is a probability and $a \in A$, such that  $\sum _{(p:a) \in \mu}p = 1$.
$\FDist(A)$ is the set of all finite multi-distributions on $A$.
For  $\mu\in\FDist(A)$, its  \emph{support}  is the multiset $\Supp(\mu)\!=\!\{a \mid (p\!:\!a)\!\in\!\mu$ for some $p\}$.

\begin{definition}[PTRS] \label{def:PTRS}
	A \emph{probabilistic rewrite rule} $\ell \to \mu \in
        \TSet{\Sigma}{ \VSet} \times \FDist(\TSet{\Sigma}{\VSet})$ is a pair  such that
 $\ell \not\in \VSet$ and 
        $\VSet(r) \subseteq \VSet(\ell)$ for every $r \in \Supp(\mu)$.
	A \emph{probabilistic TRS} (PTRS) is a finite set $\R$ of probabilistic rewrite rules.
	Similar to TRSs, the PTRS $\R$ induces a \emph{rewrite relation}
        ${\to_{\R}} \subseteq \TSet{\Sigma}{\VSet} \times \FDist(\TSet{\Sigma}{\VSet})$
        where 	$s \to_{\R} \{p_1:t_1, \ldots, p_k:t_k\}$  if there is
         a position $\pi$, a rule $\ell \to \{p_1:r_1, \ldots, p_k:r_k\} \in \R$,
          and a substitution $\sigma$
    such that $s|_{\pi}=\ell\sigma$ and $t_j = s[r_j\sigma]_{\pi}$ for all $1 \leq j \leq k$.
	We call $s \to_{\R} \mu$ an \emph{innermost} rewrite step (denoted $s \itor \mu$) if every proper subterm of the used redex $\ell\sigma$ is in normal form w.r.t.\ $\R$.
\end{definition}

\begin{example}
  \label{example:PTRS-random-walk}
  As an example, consider the PTRS $\R_{\trw}$  with the only rule $\tg(x) \to \linebreak
  \{\nicefrac{1}{2}:x, \; \nicefrac{1}{2}:\tg(\tg(x))\}$ over a signature with a unary
  symbol $\tg$ and a
  constant symbol $\O$, which corresponds to a symmetric random walk.
\end{example}

To track all possible rewrite sequences (up to non-determinism) with their corresponding probabilities, we \emph{lift} $\to_{\R}$ to \emph{rewrite sequence trees (RST)}.
An $\R$-RST is a tree whose nodes $v$ are labeled by pairs $(p_v:t_v)$ of a
probability $p_v$ and a term $t_v$. For each node $v$ with the successors $w_1,
\ldots, w_k$, the edge relation represents a probabilistic rewrite step, i.e., $t_v \itor
\{\tfrac{p_{w_1}}{p_v}:t_{w_1}, \ldots, \tfrac{p_{w_k}}{p_v}:t_{w_k}\}$. 
The root of an RST is always labeled with the probability $1$.
For an $\R$-RST $\F{T}$ we define $|\F{T}|_{\ctleaf} = \sum_{v \in \ctleaf} p_v$, where
$\ctleaf$ is the set of all leaves, and we say that a PTRS $\R$ is \emph{almost-surely
innermost terminating (iAST)} if $|\F{T}|_{\ctleaf} = 1$ holds for all
$\R$-RSTs $\F{T}$. 
While we have $|\F{T}|_{\ctleaf} = 1$ for every finite RST $\F{T}$, 
for infinite RSTs $\F{T}$ we may have  $|\F{T}|_{\ctleaf} <1$ or
even $|\F{T}|_{\ctleaf} = 0$ if  $\F{T}$ has no leaf at all.
This notion of AST is the same as the one in \cite{avanzini2020probabilistic}, 
where AST is defined using a lifting of $\to_{\R}$ to multisets instead of trees.

\begin{wrapfigure}[4]{r}{0.3\textwidth}
  \scriptsize
  \vspace*{-1.1cm}
        \begin{tikzpicture}
        \tikzstyle{adam}=[rectangle,thick,draw=black!100,fill=white!100,minimum size=4mm]
        \tikzstyle{empty}=[rectangle,thick,minimum size=4mm]
        
        \node[adam] at (-4, 0)  (a) {$1:\tg(\O)$};
        \node[adam] at (-5, -0.7)  (b) {$\nicefrac{1}{2}:\tg^2(\O)$};
        \node[adam] at (-3, -0.7)  (c) {$\nicefrac{1}{2}:\O$};
        \node[adam] at (-6, -1.4)  (d) {$\nicefrac{1}{4}:\tg^3(\O)$};
        \node[adam] at (-4, -1.4)  (e) {$\nicefrac{1}{4}:\tg(\O)$};
        \node[empty] at (-6.5, -2)  (f) {$\ldots$};
        \node[empty] at (-5.5, -2)  (g) {$\ldots$};
        \node[empty] at (-4.5, -2)  (h) {$\ldots$};
        \node[empty] at (-3.5, -2)  (i) {$\ldots$};
        
        \draw (a) edge[->] (b);
        \draw (a) edge[->] (c);
        \draw (b) edge[->] (d);
        \draw (b) edge[->] (e);
        \draw (d) edge[->] (f);
        \draw (d) edge[->] (g);
        \draw (e) edge[->] (h);
        \draw (e) edge[->] (i);
        \end{tikzpicture}
   \end{wrapfigure}  
\begin{example}
  For the infinite $\R_{\trw}$-RST $\F{T}$ on the side
  we have $|\F{T}|_{\ctleaf} = 1$.
\end{example}

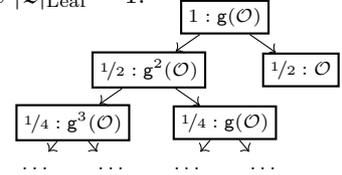
\Cref{theorem:ptrs-direct-application-poly-interpretations} introduces a novel technique
to prove AST automati\-cally by a direct application of polynomial interpretations.
The proof idea is based on \cite{mciver2017new}, but extends
it from while-programs on integers to terms.
A \emph{polynomial interpretation} $\Pol$ is a $\Sigma$-algebra with carrier
$\IN$ map\-ping every function symbol $f \in \Sigma$ to a polynomial $f_{\Pol} \in \IN[\VSet]$.
For a term $t \in \TSet{\Sigma}{ \VSet}$, $\Pol(t)$ is the interpretation of $t$ by the $\Sigma$-algebra $\Pol$.
An inequation $\Pol(t_1) > \Pol(t_2)$ \emph{holds} if it is true for all instantiations of its variables by natural numbers.

\begin{restatable}[Proving AST with Polynomial Interpretations]{theorem}{ASTPolInt}\label{theorem:ptrs-direct-application-poly-interpretations}
	Let $\R$ be a PTRS and let $\Pol:\TSet{\Sigma}{\VSet} \to \IN[\VSet]$ be a
        monotonic, multilinear\footnote{Multilinearity means that
for all $f \in \Sigma$, all monomials of $f_{\Pol}(x_1,\ldots,x_n)$ have the form $c \cdot x_1^{e_1} \cdot \ldots \cdot x_n^{e_n}$ with $c \in \IN$ and $e_1,\ldots,e_n \in \{0,1\}$. 
        As in \cite{avanzini2020probabilistic},
        multilinearity ensures ``monotonicity'' w.r.t.\ expected values, since
        multilinearity implies $f_{\Pol}(\ldots, \sum_{1 \leq j \leq k}p_j \cdot
        \Pol(r_j), \ldots) = \sum_{1 \leq j \leq k}p_j \cdot \Pol(f(\ldots, r_j,
        \ldots))$.} polynomial interpretation.
        If for every rule $\ell \to \{p_1:r_1, \ldots, p_k:r_k\} \in \R$,
    \begin{enumerate}
		\item there exists a $1 \leq j \leq k$ with $\Pol(\ell) >
                  \Pol(r_j)$ and
		\item $\Pol(\ell) \geq \sum_{1 \leq j \leq k} \; p_j \cdot \Pol(r_j)$,
	\end{enumerate}
	then $\R$ is AST.
\end{restatable}

\begin{example}\label{example:direct-application-AST}
	To prove that $\R_{\trw}$ is AST with \Cref{theorem:ptrs-direct-application-poly-interpretations}, we can use the polynomial interpretation that maps $\tg(x)$ to $x+1$ and $\O$ to $0$. 
\end{example}

\section{Dependency Tuples and Chains for Probabilistic Term Rewriting}\label{Dependency Tuples and Chains for Probabilistic Term Rewriting}

As in the non-probabilistic DP framework, we decompose the signature $\Sigma$ into defined
symbols $\Sigma_{D}$ and constructor symbols $\Sigma_{C}$. For every $f \in \Sigma_{D}$,
we introduce a fresh \emph{tuple symbol} $f^{\#}$ of the same arity.
$\Sigma^\#$ is
the set of tuple symbols and
we often write $\tF$ instead of $\tf^\#$. 
For any term $t = f(t_1,\ldots,t_n) \in \TSet{\Sigma}{\VSet}$ with $f \in \Sigma_{D}$, let
$t^{\#} = f^{\#}(t_1,\ldots,t_n)$.  
While multiple occurrences of the same subterm $f(\ldots)$
in a right-hand side of a rule
can be ignored when defining DPs for TRSs,
this is not true when analyzing PTRSs.
Hence, as in the adaption of DPs for complexity analysis in
\cite{noschinski2013analyzing}, 
we work with \emph{dependency tuples} instead of pairs.

For any $t \in \TSet{\Sigma}{\VSet}$,
if $\{t_{1}, \dots, t_{n}\}$ is the \emph{multiset} of all subterms of $t$ with defined root symbols, then we define $dp(t) = \Com{n}(t^\#_1, \ldots, t^\#_n)$. 
To make $dp(t)$ unique, we use a total order on positions.
Here, we extend $\Sigma_{C}$ by a fresh \emph{compound} constructor symbol $\Com{n}$ of arity $n$ for every $n \in \mathbb{N}$.
When rewriting a subterm $t^\#_i$ of $\Com{n}(t^\#_1, \ldots, t^\#_n)$ with a dependency tuple, one obtains terms with nested compound symbols.
To flatten nested compound symbols and to abstract from the order of their arguments, we
always 
\emph{normalize} terms implicitly.
So for example, $\Com{3}(x, x, y)$ is a normalization of $\Com{2}(\Com{1}(x), \Com{2}(x,y))$.

Instead of considering a rule $\ell \to \{ p_1:r_1, \ldots, p_k:r_k \}$ from $\R$ and its
corres\-ponding dependency tuple $\ell^\# \to \{ p_1:dp(r_1), \ldots, p_k:dp(r_k)\}$
separately, we cou\-ple them together to
$\langle \ell^\#,\ell \rangle \to \{ p_1:\langle dp(r_1),r_1 \rangle, \ldots, p_k:\langle
 dp(r_k),r_k \rangle\}$.
  So in the ``second component'', we can access the original rewrite rule used to create the dependency tuple.
 The resulting type of rewrite system is called a \emph{probabilistic pair term rewrite system (PPTRS)}.
 Our new DP framework operates on \emph{DP problems} $(\PP,\SSS)$, where
 $\PP$ is a PPTRS and $\SSS$ is a PTRS. 

\begin{definition}[Coupled Dependency Tuple] \label{def:coupled-dependency-pairs}
	Let $\R$ be a PTRS.
	For every $\ell \to  \{p_1:r_1, \ldots, p_k:r_k\} \in \R$, its 
     \emph{(coupled) dependency tuple} (\emph{DT})
        is $\langle \ell^\#,\ell \rangle \to \{ p_1 : \langle dp(r_1),r_1 \rangle, \ldots, p_k : \langle dp(r_k),r_k \rangle\}$.
	The set of all coupled dependency tuples of $\R$ is denoted by $\DTuple{\R}$.
\end{definition}

A PPTRS can  rewrite a tuple term $t^\#$ and simultaneously
use the original rewrite
rule
(that was used to create the dependency tuple) in order to rewrite all ``copies'' $t$ of
$t^\#$.
We also introduce an analogous new rewrite relation for PTRSs, where we can apply the same
rule simultaneously to the same subterms in a  single rewrite step.

\begin{example}
    Consider a PTRS $\R_{\ta}$ that (also) contains the rule $\ta \to
    \{\nicefrac{1}{2}:\ts(\tb_1), \nicefrac{1}{2}:\ts(\tb_2)\}$ and has the defined symbols $\Sigma_{D} = \{\tf, \ta, \tb_1, \tb_2\}$.
    The corresponding DT is $\langle\tA,\ta\rangle \to \{\nicefrac{1}{2}:\langle \Com{1}(\tB_1),\ts(\tb_1)\rangle, \nicefrac{1}{2}:\langle \Com{1}(\tB_2),\ts(\tb_2)\rangle\}$.
To obtain a sound termination criterion, it must be possible to mimic every
rewrite step by a corresponding chain of DTs. With our notion of PTRSs,
we can indeed mimic
 the rewrite step $\tf(\ta) \to_{\R_{\ta}} \{\nicefrac{1}{2}:\tf(\ts(\tb_1)), \nicefrac{1}{2}:\tf(\ts(\tb_2))\}$ using a single step with the PPTRS $\DTuple{\R_{\ta}}$:
    In the corresponding chain we start with  $\Com{2}(\tF(\ta), \tA)$ and
    rewrite it with $\DTuple{\R_{\ta}}$ to
$\{\nicefrac{1}{2}:\Com{2}(\tF(\ts(\tb_1)), \tB_1), \nicefrac{1}{2}:\Com{2}(\tF(\ts(\tb_2)),\tB_2)\}$.
 So here we have to access the original rewrite rule to rewrite the ``copy''
    $\ta$ of $\tA$  to $\ts(\tb_1)$ and $\ts(\tb_2)$,
    respectively.
    \end{example}

The $(\PP,\SSS)$-chains in the probabilistic setting are now defined as \emph{chain trees (CTs)}, where the edges either describe steps with the PPTRS
$\PP$ or with the PTRS $\SSS$.
Regarding the paths in this tree allows us to adapt the idea of chains, i.e., that one uses only finitely many $\SSS$-steps before the next step with a DT from $\PP$.
We define $|\F{T}|_{\ctleaf}$ and iAST for CTs as for RSTs.
With this new type of DTs and CTs, we can now mimic every possible
RST using a CT.
More precisely, from every RST we can create a CT with the same tree structure,
and for each node $v$ we have $dp(t_v^{RST}) = t_v^{CT}$, i.e., the term $t_v^{CT}$ at
node $v$ in the CT corresponds to the
$dp$ transformation of the term $t_v^{RST}$ in the RST (except for some irrelevant normal forms).
Hence, we get an analogous chain criterion to the non-probabilistic setting.

\begin{restatable}[Chain Criterion]{theorem}{ProbChainCriterion}\label{theorem:prob-chain-criterion}
    A PTRS $\R$ is iAST if $(\DTuple{\R},\R)$  is iAST.
\end{restatable}

In contrast to the non-probabilistic case, our chain criterion is \emph{sound} but not \emph{complete} (i.e., we do not have ``iff'' in \Cref{theorem:prob-chain-criterion}).
However, we also developed a refinement where our chain criterion is made complete by also storing the positions of the defined symbols in $dp(r)$.

\begin{example}
    The PTRS $\R_{\mathsf{incompl}}$ with the three rules $\tg \!\to\!  \{ \nicefrac{5}{8} : \tf(\tg), \nicefrac{3}{8} : \tstop \}$, $\tg \!\to\! \{ 1 : \tb\}$, and $\tf(\tb) \!\to\! \{ 1 : \tg\}$ shows that the chain criterion of \Cref{theorem:prob-chain-criterion} is not complete.
    \textsf{AProVE} can prove AST via \Cref{theorem:ptrs-direct-application-poly-interpretations} with
    the polynomial interpretation that maps $\tf(x)$ to $x + 2$, $\tg$ to $4$, $\tb$ to $3$, and $\tstop$ to $0$.
    On the other hand, the dependency tuples for this PTRS are $\langle \tG, \tg \rangle \!\to\!  \{ \nicefrac{5}{8} : \langle \Com{2}(\tF(\tg), \tG),
    \tf(\tg) \rangle, \nicefrac{3}{8} : \langle \Com{0}, \tstop \rangle \}$, $\langle \tG, \tg \rangle \!\to\! \{ 1 : \langle \Com{0}, \tb \rangle\}$, and $\langle \tF(\tb),  \tf(\tb) \rangle \!\to\! \{ 1 : \langle \Com{1}(\tG), \tg
    \rangle \}$.
    The DP problem $(\DTuple{\R_{\mathsf{incompl}}},\R_{\mathsf{incompl}})$ is not iAST.
    To see this, consider the following chain tree:

    \vspace*{-0.1cm}
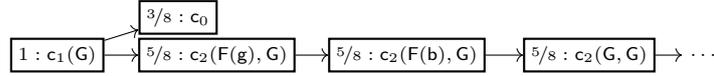
    \begin{center}
        \scriptsize
        \begin{tikzpicture}
            \tikzstyle{adam}=[rectangle,thick,draw=black!100,fill=white!100,minimum size=4mm]
            \tikzstyle{empty}=[rectangle,thick,minimum size=4mm]
            
            \node[adam] at (0, 0)  (a) {$1:\Com{1}(\tG)$};
            \node[adam] at (2.1, 0)  (b) {$\nicefrac{5}{8}:\Com{2}(\tF(\tg), \tG)$};
            \node[adam] at (1.59, 0.5)  (c) {$\nicefrac{3}{8}:\Com{0}$};
            \node[adam] at (4.6, 0)  (d) {$\nicefrac{5}{8}:\Com{2}(\tF(\tb), \tG)$};
            \node[adam] at (7.0, 0)  (e) {$\nicefrac{5}{8}:\Com{2}(\tG, \tG)$};
            \node[empty] at (8.5, 0)  (f) {$\ldots$};
            
            \draw (a) edge[->] (b);
            \draw (a) edge[->] (c);
            \draw (b) edge[->] (d);
            \draw (d) edge[->] (e);
            \draw (e) edge[->] (f);
            \end{tikzpicture}
    \end{center}
    
    \vspace*{-0.1cm}

    \noindent
    Here, we first use the first dependency tuple, then the rule $\tg \to \{ 1 : \tb\}$, and finally the third
    dependency tuple. 
    We essentially end up with a biased random walk where the number of $\tg$s is increased
    with probability $\nicefrac{5}{8}$ and decreased with probability $\nicefrac{3}{8}$.
    Hence, $(\DTuple{\R_{\mathsf{incompl}}},\R_{\mathsf{incompl}})$ is not iAST.
    The problem here is that the terms $\tG$ and $\tF(\tg)$ can both be rewritten to $\tG$. 
    When creating the dependency tuples, we lose the information that the $\tg$ inside the
    term $\tF(\tg)$
    corresponds to the second argument $\tG$ of the compound symbol $\Com{2}$.
    In order to obtain a complete chain criterion, one has to extend dependency tuples by positions. Then
    the second argument $\tG$ of the compound symbol $\Com{2}$ would be augmented 
    by the position of $\tg$ in the right-hand side of the first rule, because this rule
    was used to generate the dependency tuple. When rewriting $\tg$ by a rule from
    $\R_{\mathsf{incompl}}$, then terms like $\tG$ that belong to the same (or a lower)
    position have to be removed.
\end{example}

Our notion of DTs and chain trees is only suitable for \emph{innermost}
evaluation. To see this, consider the PTRSs $\R_1$ and $\R_2$ which both contain
$\tg \to \{ \nicefrac{1}{2}:\O, \nicefrac{1}{2}:\th(\tg) \}$, but in addition
$\R_1$ has the rule $\th(x) \to \{ 1: \tf(x,x) \}$ and $\R_2$ has the rule $\th(x) \to
\{1:\tf(x,x,x)\}$.
Note that when considering full rewriting, $\R_1$ is AST while $\R_2$ is not. 
In contrast, both $\R_1$ and $\R_2$ are iAST, since the
innermost evaluation strategy prevents  the application of the $\th$-rule to terms
containing $\tg$.
Our DP framework handles   $\R_1$ and $\R_2$ in the same way,
as both 
have the same DT $\langle \tG, \tg \rangle \to
\{ \nicefrac{1}{2}:\langle \Com{0}, \O \rangle,
\nicefrac{1}{2}:\langle \Com{2}(\tH(\tg),\tG), \th(\tg) \rangle \}$ and
a DT $\langle \tH(x), \th(x) \rangle \to
\{ 1: \langle \Com{0}, \tf(\ldots) \rangle\}$.
Even if we allowed the application of the second DT to terms of the form $\tH(\tg)$, we
would still obtain $|\F{T}|_{\ctleaf} = 1$
for every chain tree $\F{T}$.
So a DP framework to analyze ``full'' instead of innermost AST would be considerably more
involved.

\section{The Probabilistic DP Framework}\label{The Probabilistic DP Framework}

Now we introduce the probabilistic DP framework which keeps the
core ideas of the non-probabilistic framework.
So instead of applying one ordering for a PTRS directly as in
\Cref{theorem:ptrs-direct-application-poly-interpretations}, we want to 
benefit from modularity.
A \emph{DP processor} $\Proc$  is of the form $\Proc(\PP, \SSS) = \{(\PP_1,\SSS_1),
\ldots, (\PP_n,\SSS_n)\}$, where
 $\PP, \PP_1, \ldots, \PP_n$ are PPTRSs and  $\SSS, \SSS_1, \ldots, \SSS_n$ are PTRSs. 
A processor $\Proc$ is \emph{sound} if
$(\PP, \SSS)$ is iAST whenever $(\PP_i, \SSS_i)$ is iAST for all $1 \leq i \leq n$. 
It is \emph{complete} if $(\PP_i, \SSS_i)$ is iAST for all $1 \leq i \leq n$ whenever 
$(\PP, \SSS)$ is iAST. 

The (innermost)  $(\PP,\SSS)$-\emph{dependency graph} indicates which DTs from $\PP$ can rewrite to
each other using the PTRS $\SSS$.
The possibility of rewriting with $\SSS$ is not related to the probabilities.
Thus, for the dependency graph, we can use the \emph{non-probabilistic variant}
$\nonprob(\SSS) = \{\ell \to r_j \mid \ell \to \{p_1:r_1, \ldots, p_k:r_k\} \in \SSS, 1
\leq j \leq k\}$.

\begin{definition}[Dependency Graph]
  The node set of the
  \emph{$(\PP,\SSS)$-dependency graph} is $\PP$ and 
	there is an edge from $\langle \ell^\#_1,\ell_1 \rangle \to \{ p_1:\langle d_1,r_1 \rangle, \ldots,
  p_k:\langle d_k,r_k \rangle\}$ to $\langle \ell^\#_2, \ell_2 \rangle \to \ldots$ if there are substitutions
  $\sigma_1, \sigma_2$, and a
  $t^\#$ occurring in $d_j$ for some $1 \leq j \leq k$, such that $t^\# \sigma_1 \itononprobsstar
  \ell^\#_2 \sigma_2$ and both  $\ell_1^\# \sigma_1$ and $\ell_2^\#
  \sigma_2$ are in normal form w.r.t.\ $\SSS$.
\end{definition}
 
In the non-probabilistic DP framework, every step from one DP to the next in a chain
corresponds to an
edge in the dependency graph. Similarly, in the probabilistic setting,  for every
path from a node where a step with $\PP$ is used to the next such node  in a $(\PP,\SSS)$-CT,  there is a corresponding edge in the
$(\PP,\SSS)$-dependency graph. Since every infinite path in a CT contains infinitely
many such nodes, when tracking the arguments of the compound symbols,
every such path traverses a
cycle of the dependency graph infinitely often. Thus, it again suffices  
to consider the SCCs of the dependency graph separately.
To automate the following processor, the same over-approximation techniques as for the
non-probabilistic dependency graph can be used.

\begin{restatable}[Probabilistic Dependency Graph Processor]{theorem}{ProbDepGraphProc}\label{theorem:prob-DGP}
For the SCCs $\PP_1, \ldots, \PP_n$ of the
  $(\PP\!,\SSS)$-dependency graph,
  \mbox{\small $\Proc_{\mathtt{DG}}(\PP\!,\SSS)=\{(\PP_1,\SSS), \ldots,
  (\PP_n,\SSS)\}$} is sound and complete.
\end{restatable}

\begin{example}\label{example:div}
As an example, consider the following PTRS $\R_{\tdiv}$ which adapts a well-known example from \cite{arts2000termination}
to the probabilistic setting.

  \vspace*{-.4cm}
  \hspace*{-1.2cm}
  {\footnotesize
  \begin{minipage}[t]{5.3cm}
    \begin{align*}
      \tminus(x,\O) & \to \{ 1:x\}   \\
       \tminus(\ts(x),\ts(y)) & \to \{
       1:\tminus(x,y)\}    \end{align*} 
  \end{minipage}
  \hspace*{-.3cm}
  \begin{minipage}[t]{9.5cm}
    \begin{align*}
  \tdiv(\O,\ts(y)) & \to \{ 1:\O\}\\
   \tdiv(\ts(x),\ts(y)) & \to \{
    \nicefrac{1}{2}:\tdiv(\ts(x),\ts(y)),
    \nicefrac{1}{2}:\ts(\tdiv(\tminus(x,y),\ts(y))) \}   
    \end{align*}
  \end{minipage}}

  \vspace*{.2cm}

  \noindent
We get    $\DTuple{\R_{\tdiv}} = \{ \eqref{R-div-deptup-1}, \ldots,
\eqref{R-div-deptup-4}\}$:

\vspace*{-.5cm}

  {\small
 \begin{align}
          \label{R-div-deptup-1}  \langle\tM(x,\O), \tminus(x,\O)\rangle & \to \{1:\langle\Com{0},\,x\rangle\}\\
          \label{R-div-deptup-2}  \langle\tM(\ts(x),\ts(y)), \tminus(\ts(x),\ts(y))\rangle &\to \{1:\langle\Com{1}(\tM(x,y)),\,\tminus(x,y)\rangle\}\\ 
          \label{R-div-deptup-3}  \langle\tD(\O,\ts(y)), \tdiv(\O,\ts(y))\rangle &\to \{1:\langle\Com{0},\,\O\rangle\}\\
          \langle\tD(\ts(x),\ts(y)), \tdiv(\ts(x),\ts(y))\rangle
& \to \{\nicefrac{1}{2}: \langle\Com{1}(\tD(\ts(x),\ts(y))),\,
          \tdiv(\ts(x),\ts(y))\rangle, \hspace*{2cm} \nonumber
          \\
 \label{R-div-deptup-4} 
 \rlap{\hspace*{-.6cm}$\nicefrac{1}{2}:\langle\Com{2}(\tD(\tminus(x,y),\ts(y)), \tM(x,y)),\,\ts(\tdiv(\tminus(x,y),\ts(y)))\rangle\}$}
  \end{align}
  }
\end{example}

  \begin{wrapfigure}[3]{r}{0.13\textwidth}
  \scriptsize
  \vspace*{-.45cm}
    \hspace*{-.1cm}\begin{tikzpicture}
    		\node[shape=rectangle,draw=black!100, minimum size=3mm] (A) at (0,0) {$\eqref{R-div-deptup-1}$};
    		\node[shape=rectangle,draw=black!100, minimum size=3mm] (B) at (1,0) {$\eqref{R-div-deptup-2}$};
    		\node[shape=rectangle,draw=black!100, minimum size=3mm] (C) at (0,.7) {$\eqref{R-div-deptup-3}$};
    		\node[shape=rectangle,draw=black!100, minimum size=3mm] (D) at (1,.7) {$\eqref{R-div-deptup-4}$};
    	
    		\path [->,loop below,in=340,out=20,looseness=6] (B) edge (B);
    		\path [->,loop above,in=340,out=20,looseness=6] (D) edge (D);
    		\path [->] (B) edge (A);
    		\path [->] (D) edge (C);
    		\path [->] (D) edge (B);
            \path [->] (D) edge (A);                
    	\end{tikzpicture}      
  \end{wrapfigure}
The $(\DTuple{\R_{\tdiv}}, \R_{\tdiv})$-dependency graph is shown on the side. So we
obtain $\Proc_{\mathtt{DG}}(\DTuple{\R_{\tdiv}},\R_{\tdiv}) = \{(\{\eqref{R-div-deptup-2}\},\R_{\tdiv}), (\{\eqref{R-div-deptup-4}\},\R_{\tdiv})\}$.  

\medskip

For the \emph{reduction pair processor}, we use analogous
  constraints as in our new criterion for the direct application of polynomial
  interpretations to PTRSs (\cref{theorem:ptrs-direct-application-poly-interpretations}),
  but adapted to DP problems $(\PP, \SSS)$. Moreover, as in the original reduction pair
  processor, the polynomials only have to be weakly monotonic.
  For every rule in $\SSS$ or in the first components of
  $\PP$, we require that the expected value is weakly
decreasing.
The reduction pair processor then removes those
DTs $\langle \ell^\#,\ell \rangle \to \{ p_1:\langle d_1,r_1 \rangle, \ldots,p_k:\langle d_k,r_k \rangle \}$
 from $\PP$ where in addition there is
at least one term $d_j$ that is strictly decreasing.
Moreover, we can also rewrite with
the original rule 
$\ell \to \{ p_1:r_1, \ldots,p_k:r_k \}$
from the second component of the DT, provided
that it is also contained in $\SSS$. Therefore, to remove the dependency tuple, we also have to require that the rule
$\ell \to r_j$ is weakly decreasing.

\begin{restatable}[Probabilistic Reduction Pair Processor]{theorem}{ProbRPP}\label{theorem:prob-RPP}
  Let $\Pol:
\mathcal{T}(\Sigma \,\uplus$ $\Sigma^\#, \VSet)
\to \IN[\VSet]$ be a
weakly monotonic, multilinear
polynomial interpretation with
${\Com{n}}_{\Pol}(x_1,
\ldots, x_n) = x_1 + \ldots + x_n$. Let $\PP = \PP_{\geq} \uplus \PP_{>}$
with $\PP_> \neq \emptyset$ such that:
	\begin{enumerate}
		\item For every $\ell \to \{ p_1:r_1, \ldots,p_k:r_k \} \in \SSS$, we have  
		$\Pol(\ell) \geq \sum_{1 \leq j \leq k} p_j \cdot \Pol(r_j)$.
		\item For every $\langle \ell^\#,\ell \rangle \to \{ p_1:\langle
                  d_1,r_1 \rangle, \ldots,p_k:\langle d_k,r_k \rangle \} \in \PP$,\\
                we have $\Pol(\ell^\#) \geq \sum_{1 \leq j \leq k} p_j \cdot \Pol(d_j)$.   
		\item
		For every $\langle \ell^\#,\ell \rangle \to \{ p_1:\langle d_1,r_1
                \rangle, \ldots,p_k:\langle d_k,r_k \rangle \} \in \PP_{>}$,\\ 
                there exists a $1 \leq j \leq k$ with $\Pol(\ell^\#) > \Pol(d_j)$.\\
		If $\ell \to \{ p_1:r_1, \ldots,p_k:r_k \} \in \SSS$, 
                then we additionally
                have   $\Pol(\ell) \geq \Pol(r_j)$. 
	\end{enumerate}
	Then $\Proc_{\mathtt{RP}}(\PP,\SSS) = \{(\PP_{\geq},\SSS)\}$ is sound and complete.
\end{restatable}

\begin{example}
  The constraints of the reduction pair processor for the two
  DP problems from \cref{example:div} are satisfied by the
 polynomial
 interpretation which maps $\O$ to $0$, $\ts(x)$ to $2 \cdot x+1$, and
 the other function symbols to the projection on their first arguments.
This results in DP problems
 of the form $(\emptyset,
\ldots)$ and subsequently, $\Proc_{\mathtt{DG}}(\emptyset,
\ldots)$ yields $\emptyset$. By the soundness of all processors, this proves that
$\R_{\tdiv}$ is iAST. In contrast, similar to the non-probabilistic setting, a direct
application of polynomial interpretations via
\Cref{theorem:ptrs-direct-application-poly-interpretations} fails for this example. 
\end{example}

In addition to these two processors, we also have processors that can remove probabilistic
rewrite rules from $\SSS$ or even remove terms from the first components of the
right-hand sides  of DTs in $\PP$.
Furthermore, if we eventually end up with dependency tuples and rules that only have the
trivial probability 1,  we can transform the problem into a non-probabilistic DP problem and use the non-probabilistic framework.

\section{Evaluation}
We implemented our contributions in our termination prover
\textsf{AProVE},
which yields the first tool to prove almost-sure innermost termination of
PTRSs on arbitrary data structures (including 
PTRSs
that are not PAST). In our experiments, 
we compared the direct application of polynomials for proving AST (via
our new \cref{theorem:ptrs-direct-application-poly-interpretations}) with the
probabilistic DP framework. 
We evaluated \textsf{AProVE}
on a collection  of 67 PTRSs
which includes many typical probabilistic algorithms.
For example,
it contains a PTRS $\R_{\tqs}$ for probabilistic quicksort that has rules for choosing a pivot element that are only AST but not terminating, see \cite{report}.
Using the probabilistic DP framework,
\textsf{AProVE} can prove iAST of $\R_{\tqs}$ and many other typical programs.

61 of the 67 examples in our collection are iAST and \textsf{AProVE}
can prove iAST for 53 (87\%) of them.
Here, the DP framework proves iAST for 51 examples and
the direct application of polynomial interpretations via
 \cref{theorem:ptrs-direct-application-poly-interpretations}
 succeeds for 27 examples. (In contrast, 
 proving PAST via the direct application of
 polynomial interpretations as in \cite{avanzini2020probabilistic} only works for 22 examples.)
 The average
runtime of \textsf{AProVE} per example was 2.88~s (where no example took longer than 8~s).
So our experiments indicate
that the power of the DP framework
can now also be used for probabilistic TRSs.

For details on our experiments and for instructions on how to run our implementation
in \textsf{AProVE} via its \emph{web interface} or locally, we refer to \url{https://aprove-developers.github.io/ProbabilisticTermRewriting/}.



\providecommand{\noopsort}[1]{}

\end{document}